\documentclass[twocolumn,secnumarabic,amssymb, nobibnotes, aps, prd]{revtex4}
\usepackage{graphicx}

\begin{document}
\title{Could nanostructure be unspeakable quantum system?}
\author{V.V. Aristov and  A.V. Nikulov}
\affiliation{Institute of Microelectronics Technology and High Purity Materials, Russian Academy of Sciences, 142432 Chernogolovka, Moscow District, RUSSIA.} 
\begin{abstract} Heisenberg,  Bohr and others were forced to  renounce on the description of the objective reality as the aim of physics because of the paradoxical quantum phenomena observed on the atomic level.  The contemporary quantum mechanics created on the base of their positivism point of view must divide the world into speakable apparatus which amplifies microscopic events to macroscopic consequences and unspeakable quantum system. Examination of the quantum phenomena corroborates the confidence expressed by creators of quantum theory that the renunciation of realism should not apply on our everyday macroscopic world. Nanostructures may be considered for the present as a boundary of realistic description for all phenomena including the quantum one.
 \end{abstract}

\maketitle

\narrowtext

\section*{Introduction}

The progress in physics and engineering of the XX century made thanks to the quantum theory is immensely impressive. But John Bell in his famous Introductory remarks at Naples-Amalfi meeting, 1984 "Speakable and unspeakable in quantum mechanics" [1] stated that {\it "This progress is made in spite of the fundamental obscurity in quantum mechanics. Our theorists stride through that obscurity unimpeded... sleepwalking?"} Bell, as well as Einstein, Schrodinger and other opponents of the Copenhagen interpretation, connected this fundamental obscurity with the object of quantum description. The quantum mechanics studied during last eighty years was created on the base of the positivism point of view of Heisenberg and Bohr according to which {\it "any observation of atomic phenomena should include an interaction they with equipment used for the observation which can not be neglected"} [2], {\it "we had introduced an element of subjectivism into the theory, as if we meant to say: what happens depends on our way of observing it or on the fast that we observe it"} [3] and {\it "there is no way of describing what happens between two consecutive observations"} [3]. Arguing against the positivism point of view of Heisenberg, Bohr and other adherents of the Copenhagen interpretation Einstein persisted that {\it "it must seem a mistake to permit theoretical description to be directly dependent upon acts of empirical assertions, as it seems to me to be intended in Bohr's principle of complementarity"} \cite{Aristov4}. Following to Einstein Bell has located the cardinal problem of the orthodox quantum mechanics: {\it "how exactly is the world to be divided into speakable apparatus...that we can talk about...and unspeakable quantum system that we can not talk about?"} [1]. 

The positivism of Heisenberg and Bohr with its necessity to refer to 'apparatus' when atomic phenomena are discussed may be justified with the necessity to amplify microscopic events to macroscopic consequences, which we can observe. But we should call this necessity in question for macroscopic phenomena including the quantum one. There is no forcible reason to renounce on macroscopic realism and used the superposition principle contradicting to it [5] on the macroscopic level. The superposition principle was postulated for description of phenomena and it can not be interpreted as a realistic description without contradiction with locality principle, as the EPR paradox [6] has revealed as far back as 1935. The Bohr's reply [7] on the EPR critique [6] of quantum mechanics {\it "The trend of their argumentation, however, does not seem to me adequately to meet the actual situation with which we are faced in atomic physics"} witnesses that their debate on reality applied only to atomic level. Einstein was sure that {\it "in the macroscopic sphere it simply is considered certain that one must adhere to the program of a realistic description in space and time.… No one is likely to be inclined to attempt to give up this program within the realm of the "macroscopic""} [4]. In accordance to the belief of the quantum theory creators the experimental evidences of the violation of realistic prediction was observed for the present only for the atomic world, i.e. for the level of elementary particles [8]. Thus, on the one hand no experimental results force us to doubt in the reality of our everyday macroscopic world but on the other hand some phenomena observed on the atomic level, such as the double-slit interference experiments and experimental evidence of the EPR correlation [9] force us to do this. Nanotechnology comes nearer to atomic level and there is important to know where may be a boundary of these phenomena. The research for this boundary is important also because of such new developments as quantum information, quantum computation, quantum cryptography, and quantum teleportation [10], base on the EPR correlation.

\section {Two-slit interference experiment}
Richard Feynman emphasized that the double-slit interference experiment is at the heart of quantum mechanics [11]: {\it "In reality, it contains the only mystery, the basic peculiarities of all of quantum mechanics"}. Indeed, this experiment demonstrates very clear both advances and defects of universally recognized quantum formalism. The wave function formalism describes very well the interference of different particles: electrons, neutrons, atoms and even molecules [9]. But it can not explain how indivisible particles manage to pass through two slits at once or, if they do not pass, why we observe the interference pattern with period $P = L\lambda /d$ corresponding to a distance $d$ between slits, a distance $L$ between the screen with double-slits and the detecting screen and a de Broglie wavelength $\lambda = h/p = h/mv$. The de Broglie wavelength $\lambda = h/mv \approx  h/ga^{3}v$ increases with a particle sizes $a$, at the same velocity $v$ and density $g$. Recently the A. Zeilinger team [12] has observed first interference of objects with nano-sizes, biomolecules and fullerenes with length, up to  $a \approx  3 \ nm$, much larger than their de Broglie wavelength $\lambda \approx  0.004 \ nm$. A. Zeilinger in the talk "Exploring the Boundary between the Quantum and the Classical Worlds" [13] told on an intention to observe quantum interference viruses and possibly even nanobacteria. 

In order to observe an interference of a object the period of the interference pattern $P$ and distance $d$ between slits must be larger than the object size $a$: $P = L\lambda /d > a$; $d > a$. The object covers the distance $L$ between screens during a time $L/v$. Therefore the interference experiment with a object having the size $a$ should continue during the time 
$$t_{exp} > \frac{g}{h} a^{5} \eqno{(1)}$$ 
Since the density of all matters $g \approx  1000 \ kg/m^{3}$ in order of value $g/h \approx  10^{36} \ s/m^{5}$ and during $t_{exp} \approx  1 \ s$  the interference of a object with size $a < 60 \ nm$ can be observed. This size can not be increased considerably since $t_{exp} \propto  a^{5}$ (1). Thus, the level of nanostructures is boundary for a possibility of the quantum interference observation and the necessity of the superposition principle for description of such observation.

\section {EPR correlation}
In order to demonstrate the contradiction between the superposition principle and local realism, EPR [6] consider two particles states of which is entangled with a conservation law.  For example, in the Bohm's version of the EPR paradox [14] the spin states is entangled 
$$\psi =\frac{ \psi_{a,\uparrow }(r_{a}) \psi_{b,\downarrow  }(r_{b})+ \psi_{a,\downarrow }(r_{a}) \psi_{b,\uparrow }(r_{b})}{\surd 2} \eqno{(2)}$$
with the law of angular momentum conservation. Measurement of a $i$ spin projection of the particle $a$ must change instantaneously the quantum state of the both particles because of the superposition collapse 
$$\psi = \psi_{a,\uparrow }(r_{a}) \psi_{b,\downarrow  }(r_{b}) \eqno{(3)}$$
irrespective of the distance $r_{a}- r_{b}$ between their. 

The superposition principle is used for description of the outcomes of spin projection experiments because of their paradoxicality. According to the Stern-Gerlach experiment made as far back as 1922 magnetic moment $M_{1/2}$ projection $m_{1/2,i}$ of a spin 1/2 particle seems to equal the same value independently of the measurement direction $\vec{i}$. Bell has proposed in [15] a realistic description of this paradoxical phenomenon using hidden variable. But he has revealed  also in his famous theorem [16] contradictions between predictions giving with the superposition principle and any local realistic theory for the outcomes of entangled states (2) experiments. The experimental evidence [8] of Bell's inequality violation means that the EPR correlation, contradicting to the local realistic theory, is observed. But this challenge to realism can be applied only on a scale of the Bohr magneton $\mu _{B}$ and the Planck's constant $\hbar $.

In accordance to the Bohr's correspondence principle a measurement of $i$ projection gives the "classical" outcome $m \approx  (\vec{M}\vec{i}) = Mcos(\varphi )$ for a macroscopic $|\vec{M}| \gg  \mu _{B}$ magnetic moment $\vec{M}$, but no paradoxical one $m_{1/2,i} = \pm \mu _{B}$. One would think that the macroscopic quantum phenomena, superconductivity and superfluidity, violate the correspondence principle. Some authors claim on quantum superposition of macroscopic states [5,17] of molecules with magnetic moment $M \approx 200 \mu _{B}$ [18], ferrimagnetic nanoparticles with $M \approx 10^{5} \mu _{B}$ [19] and superconducting loop (SQUID) with $M \approx 10^{10} \mu _{B}$ [20]. But these claims on violation of macroscopic realism [5] have no valid experimental substantiation and are at variance with the fundamental law of angular momentum conservation and the universally recognized quantum formalism [21]. Thus, nanostructures may be considered for the present as a boundary between speakable and unspeakable quantum system.

\section*{Acknowledgement}
This work has been supported by a grant "Possible applications of new mesoscopic quantum effects for making of element basis of quantum computer, nanoelectronics and micro-system technic" of the Fundamental Research Program of ITCS department of RAS and the  Russian Foundation of Basic Research grant 08-02-99042-r-ofi.

\end{document}